\documentclass[10pt,journal,compsoc]{IEEEtran}

\ifCLASSOPTIONcompsoc
\usepackage[nocompress]{cite}

\usepackage{amsmath,amsfonts}
\usepackage{algorithmic}
\usepackage{array}
\usepackage{textcomp}
\usepackage{stfloats}
\usepackage{url}
\usepackage{verbatim}
\usepackage{graphicx}

\usepackage{balance}

\usepackage{graphicx}

\usepackage{amsthm}
\usepackage{amsmath,amssymb,amsfonts}
\usepackage{graphicx}
\usepackage{textcomp}
\usepackage{xcolor}
\usepackage{url}
\usepackage{caption}
\usepackage{subcaption}
\usepackage{multirow}
\usepackage{colortbl}
\usepackage{framed}

\newtheorem{definition}{Definition}
\newcommand{\fn}[1]{\textup{#1}}

\begin{document}
\title{An Incentive-Compatible Mechanism for Decentralized Storage Network}

\author{Iman~Vakilinia,~\IEEEmembership{Senior Member,~IEEE,}
	Weihong~Wang,
	and~Jiajun~Xin,
	\IEEEcompsocitemizethanks{\IEEEcompsocthanksitem I.Vakilinia is with the School of Computing, University of North Florida, Jacksonville,
		FL, 32224.\protect\\
		E-mail: i.vakilinia@unf.edu
		\IEEEcompsocthanksitem W.Wang and J.Xin are with The Hong Kong University of Science and Technology.}}


\IEEEtitleabstractindextext{

\begin{abstract}

The dominance of a few big companies in the storage market arising various concerns including single point of failure, privacy violation, and oligopoly.
To eliminate the dependency on such a centralized storage architecture, several Decentralized Storage Network (DSN) schemes such as Filecoin, Sia, and Storj have been introduced. 
DSNs leverage blockchain technology to create a storage platform such that the micro storage providers can also participate in the storage market. To verify the accurate data storage by the storage providers during a storage contract, DSNs apply a Proof of Storage (PoS) scheme to continuously inspect the storage service.
However, continuous verification of the storage provider imposes an extra cost to the network and therefore end-users. Moreover, DSN's PoS verification is vulnerable to a service denying attack in which the storage provider submits valid PoS to the network while denying the service to the client. 

Considering the benefits and existing challenges of DSNs, this paper introduces a novel incentive-compatible DSN scheme. 
In this scheme, the PoS is conducted only if the client submits a challenge request.  
We model the storage service as a repeated dynamic game and set the players' payoffs such that the storage provider's dominant strategy is to honestly follow the storage contract. 
Our proposed mechanism leverages the smart-contract and oracle network to govern the storage agreement between the client and storage provider efficiently.
Furthermore, our scheme is independent of a specific blockchain platform but can be plugged into any blockchain platform with smart-contract execution capability. 
As a proof of concept, we have implemented our scheme using solidity language and chainlink oracle network. The performance analysis demonstrates the applicability of our scheme.

\end{abstract}


}

\maketitle

\IEEEdisplaynontitleabstractindextext

\IEEEpeerreviewmaketitle

\IEEEraisesectionheading{\section{Introduction}}

\IEEEPARstart{N}{owadays} giant companies dominate the data storage market. The centralized architecture of such storage providers arises a number of concerns.
First, data centers are more vulnerable to the single point of failure causing the data breach, data outage, and facilitating censorship. 
Second, such companies misuse clients' personal data to earn more profit. 
Third, prices and rules are dictated by a few big players causing oligopoly. This is due to the lack of competitiveness and the small number of service providers~\cite{de2021exploring}. 

The Decentralized Storage Network (DSN) has offered a storage platform where micro storage providers can also participate in the storage market. DSNs leverage blockchain technology to facilitate the management of the storage service. 
Blockchain applies the distributed ledger to store transaction histories, and the information is stored across a network of computers instead of on a single server. 
Utilizing the smart contract, various incentivization mechanisms can be developed on top of the blockchain to automatically moves digital assets following arbitrary pre-specified rules. 

DSNs provide an algorithmic storage market to clients and storage providers. The client can outsource the data storage by making a payment to the network. On the other hand, the storage providers share their storage resources with the network in return for a premium.  
To satisfy data confidentiality, clients' data is encrypted end-to-end at the client side and storage providers do not have access to the decryption keys. This can offer enhanced security and privacy by eliminating the central entity that controls the data. 
Moreover, micro storage providers can rent out their excessive storage resources, improve the network throughput and reduce the maintenance cost of data centers. 
As a result, the storage service can be delivered cheaper with more players and options. 

One of the main challenges in the decentralized storage network is to verify the correct storage of data by the storage provider. To this end, a DSN platform should be equipped with a \textit{Proof of Storage} (PoS) scheme to monitor the honest behavior of the storage provider for storing the outsourced data intact~\cite{ateniese2020proof}. 
Currently, DSNs perform PoS periodically during the storage contract to ensure the accurate storage of the data by the storage provider~\cite{filecoin,sia,storj}. 
However, such a continuous verification is costly and it is vulnerable to a malicious storage provider that submits PoS to the DSNs nodes while refusing service to the client.  

To improve the DSN's performance and protect the client from service denying attack, in this paper, we present and analyze a novel decentralized storage scheme in which it is not required to verify the storage service constantly but only once challenged. 
To this end, we design a new \textit{Incentive-Compatible} mechanism such that players achieve their best outcomes by choosing their actions truthfully. More specifically, we design a repeated dynamic game such that the storage provider's strategy for honestly sharing the stored data is the only subgame-perfect equilibrium. We utilize the smart contract and oracle network to enforce the rules of our proposed storage game. 
Eliminating the continuous storage verification significantly improves the performance of the DSN. On the other hand, our incentive design supports clients from the service denying attack as we explain further in next sections.

Our proposed scheme has the following benefits compared with the existing DSNs:

\begin{itemize}
	\item Current DSNs require continuous proof from the storage provider to ensure that the data is stored accurately. This process is costly and causes a waste of energy. However, in our scheme, the storage provider does not need to continuously prove the correct storage but only when the client challenges the storage provider.
	
	\item Current DSNs require adding third-party services or a new blockchain platform to intervene in the examination of honest behavior of client and storage providers. However, our proposed mechanism does not require a new service but it is pluggable on any blockchain with smart-contract execution capability (e.g., Ethereum). 
	
	\item The proof of storage in the current DSNs requires storage providers to proof the storage to the DSNs' network nodes. This can cause a service denying attack such that a dishonest storage provider only provides proof to the network nodes while rejecting service to the client. This threat is not credible in our scheme as we discuss later on.
	
	\item Current DSNs are unable to manage the number of data requests from clients. As a result, a dynamic pricing model such as pay-as-you-go cannot be implemented. We discuss how our proposed model can manage the number of requests to support diverse storage services based on the volume of retrieval requests.
	
\end{itemize}

To achieve these goals, first we model a storage contract as a repeated dynamic game. Then, we set the players' payoffs such that the dominant strategy of the storage provider is to provide the storage service honestly. Finally, we implement our proposed scheme using a smart-contract and oracle network.
To the best of our knowledge, this work is the first to investigate an incentive-compatible challenge-based decentralized storage mechanism utilizing the smart-contract and oracle network.

The rest of the paper is organized as follows. The next section reviews major works in the field of the decentralized storage network. In section~\ref{sys}, we overview the decentralized storage network's components. Details of our proposed mechanisms are described in section~\ref{mech}. The experiment results have been discussed in section~\ref{ana}. Finally, we conclude our paper in Section~\ref{con}.

\section{Related Work}\label{rel}

\subsection{Decentralized Storage Network}

Filecoin~\cite{filecoin}, sia~\cite{sia}, storj~\cite{storj}, and swarm~\cite{ethersphere2016sw3} are the most well-known platforms utilizing the blockchain technology to implement the DSN. 
Such platforms leverage the blockchain asset management capabilities to enforce incentive models for clients and storage providers.

Filecoin~\cite{filecoin} runs on a blockchain with a native protocol token (also called ``Filecoin'') which miners earn by providing storage to clients. Clients spend Filecoin hiring miners to store or distribute data. Filecoin miners compete to mine blocks with sizable rewards, but Filecoin mining power is proportional to active storage, which directly provides a useful service to clients.  
To earn Filecoin, storage providers must prove they are storing the data properly. The Filecoin network verifies that data is stored securely through cryptographic proofs. Storage providers submit their storage proofs in new blocks to the network and validate new blocks sent from the network. 
Filecoin applies the Proof-of-Spacetime, where a verifier can check if a prover is storing the outsourced data for a range of time.
Filecoin works as an incentive layer on top of the Interplanetary File System (IPFS). 
IPFS~\cite{ipfs} is a p2p storage network. Content is accessible through peers located anywhere in the world. These nodes relay information, store it, or do both. IPFS uses content addressing rather than location based addressing to find data. A content identifier, or CID, is a label used to point to material in IPFS. It doesn't indicate where the content is stored, but it forms a kind of address based on the content itself. 

Storj~\cite{storj} is another well-known DSN. Stoj utilizes a service called satellite to manage the decentralized storage system. Satellites are responsible for verifying storage, data repair service, receiving and distributing payments, managing storage nodes, account management and authorization system, and storing storage metadata. Storj extends the probabilistic nature of common per-file proofs-of-retrievability to range across all possible files stored by a specific node. 
Figueiredo \textit{et al}.~\cite{de2021exploring} have investigated the security of the Storj network and explored a DoS vulnerability within Storj’s dev./test environment which was experimentally evaluated to be highly feasible. The attack results in the inability of developers to access their test data and storage providers missing out on their payments. However, the authors pointed out that Storj’s production system is not vulnerable to such an attack as long as multiple satellites running on load-balanced clusters of servers.

Sia~\cite{sia} is also a famous DSN. Sia runs its own blockchain. Sia's blockchain stores the file contract. This contract includes the terms of the storage agreement such as pricing and uptime commitment. 
Sia divides files into 30 segments and uploads each segment in different nodes. This distribution assures that no one host represents a single point of failure and reinforces overall network uptime and redundancy.

Swarm~\cite{ethersphere2016sw3} is a distributed storage platform and content distribution service. The primary objective of Swarm is to provide a decentralized and redundant store of Ethereum's public record, in particular, to store and distribute decentralized applications' code and data as well as blockchain data. From an economic point of view, it allows participants to efficiently pool their storage and bandwidth resources in order to provide these services to all participants. The goal is a peer-to-peer serverless hosting, storage, and serving solution that is DDoS-resistant, has zero downtime, is fault-tolerant and censorship-resistant as well as self-sustaining due to a built-in incentive system. 
 

\subsection{Proof of Storage}
 
 One of the main challenges for having a robust DSN is to audit that the storage provider is honestly storing data intact. To this end, a Proof of Storage (PoS) scheme is used. A similar notion of PoS is Proof of Data Possession (PDP).
 If the storage provider fails to provide the proof, then it will be penalized by the DSN. 
 
 PoS schemes have been studied widely in the literature for centralized setting~\cite{dziembowski2015proofs,ateniese2007provable, ateniese2008scalable, erway2015dynamic}. However, such schemes are not necessarily applicable in the DSNs as in the decentralized setting the verification must be cheap, and the proof and public parameters must be succinct. 
 
 Filecoin has introduced a new proof scheme, called Proof of Spacetime (PoSt), where a verifier can check if a prover has indeed stored the outsourced data they committed to over space (i.e., storage) and over a period of time. In PoSt, the prover generates sequential PoS and recursively composes the executions to generate a short proof.
 Filecoin's PoSt applies zk-SNARKs~\cite{ben2013snarks} to generate succinct proofs which are short and easy to verify~~\cite{filecoin}.
 
 Storj~\cite{storj} introduces the satellite component as a third party to audit the storage service. However, using such a third party service weakens the decentralized architecture. 
 Moreover, storj utilizes the reputation based system for storage providers based on the history of their service.

 Sia~\cite{sia} uses a Merkle tree based auditing scheme such that the host is required to demonstrate the possession of a random segment. A storage provider must present a certain number of proofs to the network within the time frames specified in the file contract to get fully paid.

 Recently, several research studies have investigated new methods to improve the proof of storage for DSNs~\cite{ateniese2020proof,campanelli2020incrementally,du2021enabling}. Du \textit{et al}.~\cite{du2021enabling} have proposed a new framework for auditing the data in the DSNs using pairing based cryptography and zero-knowledge proof. Yu \textit{et al}.~\cite{yu2021efficient} have designed a data-time sampling strategy that randomly checks the integrity of multiple files at each time slot with high checking probability. Furthermore, this research proposes a fair sampling strategy by designing an arbitration algorithm with a verifiable random function. 
 
 Besides, Vector Commitments (VCs) can also serve as a PoS solution. 
 VCs can commit to a list of values to a digest, and later provide succinct proof to prove one value is the committed value in some specific location.
 However,  vector commitment has different limitations.
 RSA based VCs require a trusted setup for the hidden order group \cite{catalano2013vector, lai2019subvector, boneh2019batching, campanelli2020incrementally}.
 A class group \cite{buchmann1988key} can be used to generate the hidden order group instead of the trusted setup, but it is still not practical due to calculation overheads.
 Bilinear groups based VCs \cite{catalano2013vector, libert2010concise, gorbunov2020pointproofs, srinivasan2021hyperproofs} require at least a linear number of public parameters as common reference strings which limits its adoption in decentralized settings. 
 Lattice based VCs \cite{papamanthou2013streaming} have the pros of post-quantum security, simple setup, and cons of larger communication as well as computation overhead due to the lattice itself.

 A relative notion is Proof of Retrievability (PoR) \cite{shacham2008compact, juels2007pors, bowers2009proofs, cash2017dynamic, dodis2009proofs, stefanov2012iris, shi2013practical, fisch2018scaling}. PoR guarantees that only if a server stores entire files without loss, it can provide a valid proof. 
 While in PoS, the server can still provide a valid PoS proof with some part of the file lost with non-negligible probability.
 However, the stronger guarantees come with a price. Most PoR schemes require heavy cryptography tools, assumptions, or large overhead.

 \subsection{Motivation}

 Inspired by the previous schemes, in this paper we present a novel game-theoretic challenge-based storage contract mechanism for DSNs.
 To this end, our proposed mechanism allows the client to submit a challenge request indicating that the storage provider has not shared the outsourced data. Once the challenge request is received, smart-contract and oracle network conduct the storage verification. 
 In contrast with previous works, our scheme is designed such that it does not require a continuous verification of storage but it only executes once the client submits a challenge request. The mechanism is designed such that the dominant strategy for the storage provider is to honestly store and share data with the client. On the other hand, the client's dominant strategy is to not submit a challenge request if the storage service has been delivered accurately. 
 Moreover, our scheme prevents the dishonest storage provider to deliver PoS to the DSN while refusing the storage service to the client. 
 This design significantly decreases verification overhead costs in the current DSNs. We leverage the oracle network to alleviate the DSN's execution cost.
 Furthermore, our proposed scheme is independent of an underlying blockchain layer and can be executed on top of every generic blockchain with smart-contract execution capabilities such as Ethereum and Bitcoin. 
 Our scheme leverages the Merkle tree for the PoS as we describe in section~\ref{mech}. 

\section{Overview}\label{sys}

In this section, we review the system architecture of the storage service in a decentralized storage network utilizing blockchain technology.

A DSN provides a platform for a storage provider to offer the storage service to clients. A client aims to purchase the storage service to store and access her data for a specified time period. On the other hand, a storage provider aims to sell his storage service to host the client's data in return for a premium. 
In a nutshell, a DSN is equipped with two main components of payment settlement and storage verification. In the payment settlement module, the DSN charges the client for the storage service and makes the payment to the storage provider. Moreover, in case the storage provider fails to provide the committed storage service, the DSN penalizes the storage provider and compensates the client accordingly. 
In the verification module, the DSN verifies that the storage provider is delivering the storage service accurately. To this end, the storage provider should submit proof of storage to the DSN, and DSN verifies the correctness of such proofs.

Once the client and the storage provider agree on a storage service, they enroll in a storage contract. This contract conveys the storage Service Level Agreement (SLA) which specifies the storage service including the duration of the contract, premium, quality of service, and compensation rates. 

DSNs leverage public blockchain technology to enforce storage contracts.
Blockchain technology has offered an agreeable platform for parties to make payments without a single trusted third party. Blockchains are managed by a peer-to-peer network to manage a digital ledger. Recorded data on a public blockchain is publicly accessible and tamper-resistant. A smart contract is a code in the blockchain that automatically enforces a contract between two parties without any help from a single third party. Therefore, there is no need for an intermediary between contracting entities to enforce the contract.
Accordingly, in a blockchain enabled DSN scheme, there is not a single party controlling any storage contract.

In a DSN storage service, the client may encrypt her data before submitting it to the storage provider to protect the confidentiality of her data. Moreover, DSNs can provide redundancy, high-availability, and fail-over by storing the data in multiple nodes in the network.

Note that the details of storage techniques that the storage provider is using to store the client's data are out of the scope of this paper. In other words, we assume the storage provider manages his storage resources including redundancy, server location, backup services, network bandwidth, etc to maximize his payoff following the SLA. 

\subsection{Design Goal}

The primary goal of DSN's mechanism design is to ensure that the storage provider stores the client's data and returns it upon the client's request following the SLA. 
The storage service should be examined, and the client should be compensated in case of a storage failure. The client should pay the storage provider if the storage service has been delivered flawlessly. 
We aim to improve the current methods of PoS in DSNs by eliminating the requirement of continuous verification of the data storage on the storage provider. 
The mechanism should be incentive-compatible such that the players can earn their best outcome by choosing their actions truthfully.

Moreover, we consider the following side features in our design goal:

\begin{itemize}
	
	\item \textbf{Blockchain platform independent.} Current DSN systems work on their own customized blockchain platforms. This causes an extra overhead cost for the DSN. We aim to propose a compatible storage scheme that is pluggable to available generic blockchain platforms with smart-contract execution compatibility (e.g., Ethereum).
	
	\item \textbf{Prevent service denying attack.} Currently available DSNs are vulnerable to a service denying attack such that a dishonest storage provider denies providing the expected service to the client while successfully submitting PoS to the DSN network. In this case, the storage provider receives the service fee while the client has not received the expected service. We aim to protect the DSN network from such a fraudulent storage provider.
	
	\item \textbf{On-chain efficiency.} On-chain storage and computation are costly. Therefore, the proposed scheme should minimize the on-chain storage and computation without compromising security expectations. 
	
	\item \textbf{Counting requests.} Many storage services expect to count the number of requests from the client to dynamically calculate the cost of service. So far, the available DSNs do not provide storage services based on the number of requests. We will discuss how our proposed mechanism can accomplish this task.

\end{itemize}

\section{An Incentive-Compatible Mechanism for the Storage Contract}\label{mech}

In this section, we discuss our proposed mechanism for an incentive-compatible decentralized storage system based on smart-contract. First, we model and analyze the storage contract as a repeated dynamic game, and then we discuss the design of the storage contract utilizing the smart-contract and oracle network.

\subsection{Storage Contract As a Repeated Dynamic Game}

The mechanism design objective is to place a set of rules for the storage service to meet the requirements. 
A mechanism can be specified by a game $g: \mathcal{M} \rightarrow \mathcal{X} $ where $\mathcal{M}$ is the set of possible input messages and $\mathcal{X}$ is the set of possible outputs of the mechanism.
In the storage system model, players are the storage provider and the client. A rational player chooses his strategy to increase his/her utility. We assume players are rational self-interested such that they aim to maximize their profit. 

In the design of a decentralized storage network, the following questions need to be answered:

\begin{itemize}
	\item How the mechanism can verify the storage provider's honest behavior of sharing data?
	\item What is the payment channel for the service?
	\item How the mechanism can charge the client for the service? 
	\item How the mechanism can penalize the storage provider for loss of data or low quality service?
\end{itemize}

A naive model is to have a Trusted Third Party (TTP) mediate between the client and storage provider. In this case, the client requests data from TTP, and TTP receives data from the storage provider. TTP can check the integrity of data by storing the hash of data and verifying it whenever receiving data from the storage provider, and then forward it to the client.
Upon successful execution of the service, TTP charges the client and pays the storage provider. On the other hand, if the storage provider fails to provide data back to the client, TTP charges the storage provider and pays the client for the data loss or low quality service according to the contract. Although this model ensures the storage service expectations, it is inefficient, expensive, and not scalable due to the requirement of having a TTP as a middleman for every request and response. On the other hand, finding such a TTP is impractical, and such a design resembles a centralized architecture where the TTP acts as a central party and can be potentially bribed. 

To solve this problem, DSNs rely on a blockchain platform to act as a TTP to remove a single central entity to manage the system. On the other hand, DSNs minimize TTP intervention in the data recovery process such that the client retrieves data from the storage provider directly. However, the DSN continuously verifies the proof of storage from the storage provider to ensure the storage provider is storing the data truthfully. DSNs utilize blockchain asset management features to deliver the fees among players. Although this method notably improves the naive solution mentioned above, there are two main issues that remain:
\begin{itemize}
	\item First, the continuous verification of storage is costly for the network. 
	\item Second, a dishonest storage provider can successfully submit the proof of storage to the DSNs while refusing the service to the client.
\end{itemize}

To solve these issues, in our model the DSN does not continuously verify the storage service but whenever the client submits a challenge request. 
In our model, the client directly requests data from the storage provider, and the storage provider sends back data directly to the client. However, as a dishonest storage provider might refuse to provide data back or send back incorrect data, the mechanism is equipped with a challenging option. In this case, the client can send a challenge request to the TTP, and TTP verifies the data that the storage provider returns back.  
Therefore, in our mechanism, the interaction between the client and the storage provider can be modeled as a \textit{repeated dynamic game}.
Once the storage contract starts, in the first stage of the game, the storage provider can choose between \textit{Sharing} data or \textit{Not Sharing} data strategies. 
Here, sharing means that the storage provider honestly follows the storage contract and shares the client's data upon the client's data request. On the other hand, not sharing, indicates that the storage provider refuses the service to return data upon the client's request.  

Afterward, the client can choose \textit{Challenging} or \textit{Not Challenging} the storage provider. Challenging means that the client submits a challenge request to the DSN indicating that the storage provider has not shared data. Not challenging indicates that the client does not submit a challenge request. 
Upon receiving the challenge request, DSN performs the storage verification. 

\begin{figure}[h]
	\begin{center}
		\includegraphics[width=3.5in,height=1.9in,keepaspectratio]{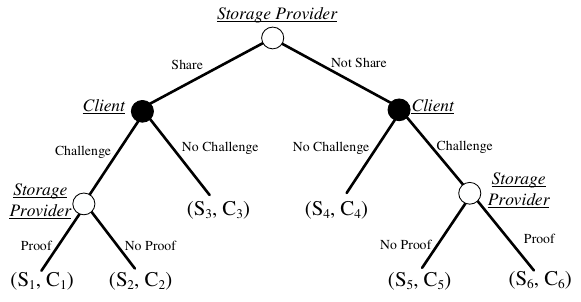} 
		\caption{Dynamic Game of the storage contract}
		\label{fig:dyngam}
	\end{center}
\end{figure}

Once the storage provider is challenged, then his strategy set is to \textit{Proof} of storage or \textit{Not Proof} of storage. Proof means that the storage provider submits the proof of storage to the DSN. Not Proof means that the storage provider does not submit the proof of storage or fails to submit the accurate proof of storage to the DSN.
This repeated dynamic game is depicted in Fig~\ref{fig:dyngam}.

The goal of our mechanism design is to ensure that the \textit{subgame-perfect-equilibrium} of this dynamic game is \textit{\{Share, No Challenge\}}. In other words, we aim to design a mechanism in which storing and sharing data is the storage provider's dominant strategy, and the client's dominant strategy is to not submit a challenge request.
To achieve this goal, first, we investigate the players' payoffs in the leaf nodes, and then we set payoffs such that the \textit{Sharing} and \textit{No Challenge} strategies are the dominant strategies for the storage provider and the client, respectively. 

As can be seen in Fig~\ref{fig:dyngam}, there are six possible outcomes for our storage contract game. 

At the first stage of the game, the storage provider's action set is \{\textit{Share}, \textit{No Share}\}. The goal of our mechanism is to ensure that the \textit{Sharing} action is the best strategy for the storage provider. However, Sharing is costly because of the cost of storage, data retrieval, backup, and network bandwidth required for successful sharing. 

In the second stage, the client action set is \{\textit{Challenge}, \textit{No Challenge}\}. By choosing the \textit{Challenge} strategy, the client claims that the storage provider has not shared data. 
If the client chooses to challenge, then the storage provider has two options as \{\textit{Proof}, \textit{Not Proof}\}. If the storage provider chooses \textit{Proof}, then it should provide proof of storage, otherwise, if the storage provider fails to proof, then the storage provider will be penalized according to the contract. 

We assume the data has a value for the client. In other words, the client receives compensation in return for the data loss caused by the storage provider. 
Note that, failure of proof is the worst outcome for the storage provider as the storage provider will be penalized the compensation amount, and we have:

\begin{equation}
	S_{i \in \{1,3,4,6\}} >> S_{j \in \{2,5\}}  
\end{equation}

Following the \textit{backward induction}, the storage provider would choose proof action unless he lost data or cannot provide the service. This is because the cost of \textit{No Proof} is the cost of compensation for the client, and we have:

\begin{equation}
	S_1 >> S_2 \quad,\quad S_6 >> S_5  
\end{equation}

As the mechanism's goal is to ensure that the storage provider chooses the \textit{Sharing} strategy, the cost of \textit{Proof} should be higher than the cost of \textit{Sharing}.
Therefore, we have:

\begin{equation}
	S_3 >> S_1 \quad,\quad S_4 >> S_6  
\end{equation}

On the other hand, as the mechanism's goal is to ensure that the client chooses \textit{No challenge} when the storage provider honestly shared data, then we should have the following:

\begin{equation}
	C_3 > C_1
\end{equation}

To this end, the mechanism makes the \textit{Challenging} request costly for the client. Let $\mathcal{X}$ represent this cost. 

On the other side, as we want the storage provider chooses \textit{Sharing}, the client should choose \textit{Challenge} when the data has not been shared by the storage provider. To achieve this goal, the mechanism should motivate the client for choosing the challenge once the data has not been shared. However, the challenge is costly as we discussed earlier. 
To cover the cost of the challenge, our mechanism is designed such that the \textit{Proof} strategy enforces a copy of data to be sent out to the client to improve the payoff of the challenge strategy in case of not sharing. On the other hand, once the data has not been shared, there is a possibility that the storage provider cannot provide the storage service (e.g., due to the data loss). Let $\mathcal{P}$ represent the probability that the storage provider cannot present the storage service. 
Let $\mathcal{V}$ represent the value of accessing data for the client. Then, the client's expected utility for choosing the challenge can be modeled as:

\begin{equation}
	C_c =  \mathcal{P}.(C_5) + (1-\mathcal{P}) . (\mathcal{V}) - \mathcal{X}
\end{equation}

Note that here $C_4<0$ as the client has not received the service. On the other hand the mechanism sets $C_c>0$, therefore we have $C_c > C_4$. Using the backward induction, it can be seen that the subgame-perfect-equilibrium of this game is \textit{\{Share, No Challenge\}} strategy profile.

\noindent\textbf{\textit{Example}}

In this section, we provide an example to clarify the proposed mechanism. For simplicity, we consider the utility of players as a number without the declaration of a specific currency.

\begin{figure}[h]
	\begin{center}
		\includegraphics[width=3.5in,height=1.9in,keepaspectratio]{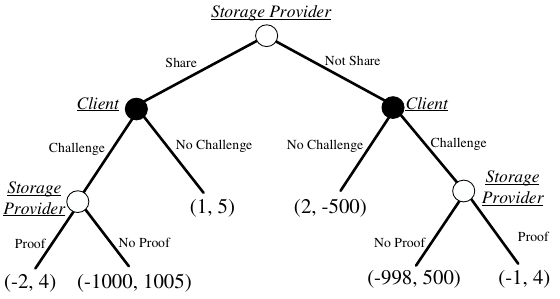} 
		\caption{Dynamic Game of the example storage contract}
		\label{fig:exam}
	\end{center}
\end{figure}

Consider that the compensation cost is indicated as ``$1,000$" in the storage contract. In other words, if the storage provider is unable to retrieve the client's data, then the storage provider should pay the client ``$1,000$". Let the cost of losing data for the client be ``$500$", and the client's benefit of reading data is ``$5$". The cost of requesting the challenge is ``$1$", and the cost of proof of storage for the storage provider is ``$3$". Finally, let the benefit of not sharing data with the client is ``$2$", and the benefit of sharing data be ``$1$" for the storage provider (note that this is the payoff that the storage provider earns by charging the client for providing the correct storage service). Therefore, the dynamic game tree of this game can be depicted as Fig~\ref{fig:exam}.

Using the backward induction, it can be seen that the \{Share, No Challenge\} is the subgame nash equilibrium of the game. 

\subsection{Scheme details}

Identifying the storage contract requirements and the players' payoffs, now we discuss the design architecture to satisfy the design goal. 

We utilize the smart-contract to act as a TTP to manage the agreement between the client and the storage provider. 
Smart-contract is powered by blockchain technology. The blockchain is managed by a peer-to-peer network to manage a digital ledger. A smart-contract is a code in the blockchain that automatically enforces a contract between two parties without any help from a third party. Therefore, there is no need for an intermediary between contracting entities to enforce the contract.
A public blockchain network capable of executing the smart-contract is used as a platform for developing the DSN. 

In our proposed model, first the client and storage provider reach a storage agreement. The agreement includes the following information:
Length of contract, Merkle root of data, premium, delivery time, and compensation. 
For example, a storage provider makes an agreement with a client to store her 1 TB data for the length of 1 year for the premium of \$20, if the storage provider fails to return data, then the storage provider will be penalized by \$40, and the time window for delivering data after client's request is 20 minutes.   

This agreement will be specified in the smart-contract and deployed on the blockchain. 
Note that, the Merkle root of data is stored on-chain which will be used for verification of data. Storing the whole data on the blockchain is too costly, therefore Merkle-tree is used to minimize the cost of the storage verification process as we discuss later on.
Moreover, the smart-contract includes the client's premium as well as the storage provider's collateral asset. Upon the successful storage service, the premium will be automatically transferred to the storage provider. On the other hand, if the storage provider fails to provide the storage service, the client will be automatically compensated through the collateral asset of the storage provider following the contract details.

Note that the blockchain platform is an isolated network, and it cannot pull in or push data out to any external system. This problem is known as the oracle problem~\cite{oracle,egberts2017oracle}. To solve this problem, the oracle network has been presented. Oracle network provides a trusted source for accessing off-chain data to the blockchain. Moreover, it can perform arbitrary programs more efficiently compared with the smart-contract, due to the fact that fewer resources are needed to execute the code. 
Leveraging the oracle network, the challenge request in our scheme works as follows:

\textit{The client submits a challenge request by calling the challenge function of the smart-contract with the specific data segment number to be challenged. 
Once the challenge request has been received by the smart-contract, the oracle network will submit a challenge to the storage provider. Upon receiving the challenge request from the oracle network, the storage provider should send the challenged data along with the Merkle path to the oracle. In the next step, the oracle network first calculates the Merkle root and compares it with the Merkle root in the storage contract stored on-chain. If they match, then the oracle sends a copy of the data to the client. Otherwise, the oracle sends a fail signal to the smart-contract, and the smart-contract will transfer the compensation fund from the storage provider account to the client account following the pre-specified agreement. }

\begin{figure}[h]
	\begin{center}
		\captionsetup{justification=centering}
		\includegraphics[width=3.5in,height=1.9in,keepaspectratio]{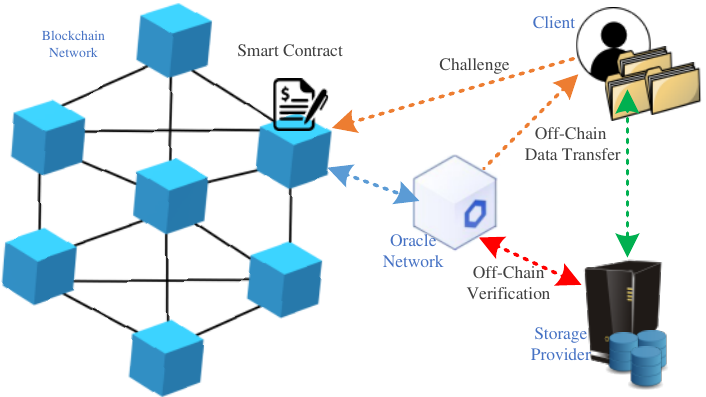} 
		\caption{Interaction of different components in the proposed challenge based DSN storage contract}
		\label{fig:scheme}
	\end{center}
	\vspace{-.2in}
\end{figure}

The interaction between different components of our scheme is depicted in Fig~\ref{fig:scheme}.

As can be seen, the data transfer is done off-chain. The only on-chain operation is the challenge request. 
Note that in this scheme, the storage provider should send the challenged data to the oracle network, and the oracle network forwards the data back to the client. There are two main reasons for this design. 

\begin{itemize}
	\item First, it prevents the service denying attack which we explained earlier. This is due to the fact that if the storage provider refuses service to the client, the client receives a copy of data with the challenge request if the storage provider can pass the proof. Therefore, the storage provider cannot deny service to the client while proofing the storage to the DSN.
	
	\item Second, when the storage provider has not shared data, the mechanism should provide incentives for the client to submit the challenge. By forwarding the data, we add the value to the client's payoff for choosing the challenge strategy to achieve our desired subgame perfect equilibrium as we discussed in the previous section. 
	
\end{itemize}

\noindent\textbf{\textit{Challenge Level}} 

To improve the efficiency of the scheme, we consider different levels for the challenge request. Let us motivate this feature by an example. Assume the client outsourced a very large dataset of 100 TB to a storage provider. If there is no challenge level option for the system, the storage provider should forward the whole data back to the client upon the challenge request submitted by the client. This can be a resource consuming task. To solve this problem, we consider that the client and storage provider agree to split data into a specific number of segments. In the challenging phase, the client can submit challenges for a set of segments. The price of challenge requests is an increasing function of the size and number of the data segments being challenged. 
For example, assume the outsourced data of size 100 TB is divided into 100,000 segments of 1 GB size. The client can submit challenges for any number of segments, however, submitting a challenge for larger data is more costly for the client to deter a malicious client to cause a denial of service attack on the storage provider. We explain how our proof of storage scheme can handle this feature in the next section.

\subsubsection{Proof of Storage}

\textit{Proof of Storage} (PoS) scheme allows a verifier to check if a storage provider is storing the client's data at the time of the challenge.
We follow the definitions from \cite{ateniese2007provable} with minor modifications.

\begin{definition}
	Given security parameter $\lambda$, the PoS scheme is a tuple of three probabilistic polynomial-time (PPT) algorithms \fn{(Setup, Prove, Verify)}:
	\begin{itemize}
		\item $(d, h) \gets \fn{Setup}(1^{\lambda}, \mathcal{D}, sz)$. This algorithm takes as input the security parameter $\lambda$, outsourced data $\mathcal{D}$, and segment size $sz$. The algorithm outputs a digest of data $d$ which is used to verify the proof and the Merkle tree height $h$.
		\item $\pi \gets \fn{Prove}(\mathcal{D}, c)$. This algorithm takes as input the data $\mathcal{D}$ and the challenge number $c$. Here, $c$ is the identifier of the challenged data.
		The algorithm outputs a proof $\pi$ to prove the storage of data segments corresponding to the challenged node number $c$.
		\item $0 / 1 \gets \fn{Verify}(d, n, c, \pi)$. This algorithm takes as input the digest $d$, the Merkle tree height $h$, the challenge number $c$, and the proof $\pi$.
		It outputs $1$ if the proof $\pi$ is a valid proof, and it outputs $0$ otherwise.
		
	\end{itemize}
\end{definition}

\noindent\textbf{\textit{Cryptographic building blocks}}

\noindent Cryptographic secure hash function. We use a cryptographic secure hash function $H \gets \fn{Hash}(x)$ that is a collision and pre-image resistant.

\noindent Digital signature. We use the standard EU-CMA secure digital signature function \cite{katz2020introduction} that contains three functions:
1) $(pk, sk) \gets \fn{KeyGen:}(1^{\lambda})$;
2) $\sigma \gets \fn{Sign}(sk,m)$;
3) $1/0 \gets \fn{Verify}(pk,m,\sigma)$.

\noindent\textbf{\textit{Our construction}}

Our construction is based on the Merkle tree. 
First, the data $\mathcal{D}$ is divided into a set of segments $\mathcal{S}=\{s_1, \dots, s_m\}$ where each segment has the size of $sz$. Here, $m$ should be the power of 2. Segments with 0 bits are padded to meet this requirement.
The hash of each segment builds the leaves of the Merkle tree. 
The parent node is the hash of its child nodes.

The Merkle tree root is labeled with the number 0, its two children nodes with the number 1 and 2. All the other nodes' numbers follow the same incremental manner. Following this numbering, the last node number is $2m-2$, and we have $h = log (m)$.

The \fn{Prove} function outputs the sibling nodes of the challenged node's Merkle path and the corresponding segments needed to calculate the challenged node.
the \fn{Verify} function checks if the node generated by the segments has a valid Merkle path towards the Merkle root.

\begin{figure}[h]
	\begin{center}
		\includegraphics[width=2.5in,height=1.5in,keepaspectratio]{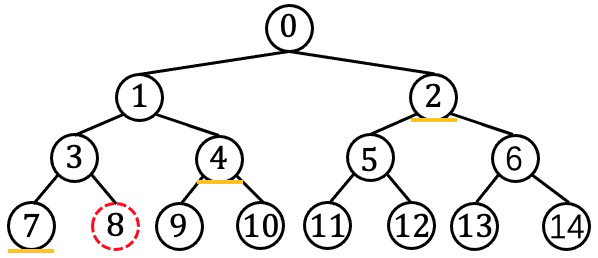} 
		\caption{Merkle tree example.}
		\label{fig:merkle}
	\end{center}
\end{figure}

For example, assume a Merkle tree as depicted in Figure \ref{fig:merkle}. 
Here, the challenged number is 8. The proof $\pi$ includes the raw data of segment $s_2$ to generate leaf node 8, and the sibling nodes of the Merkle path for node 8, which includes node numbers 2, 4, and 7. 
In order to verify the proof $\pi$, the verifier first calculates the leaf node based on the data of segment $s_2$ received from the prover. Then, it calculates node 3 using nodes 7 and 8, calculates node 1 using nodes 3 and 4, and calculates node 0 using nodes 1 and 2. Finally, it checks if the calculated root is the same as the Merkle root stored on-chain.

The process of storing the Merkle root of data on-chain is as follows. First, the client and storage provider generate their public/private key pairs. 
Let $(pk_{c},sk_{c})$ and $(pk_{sp},sk_{sp})$ to denote their public/private key pairs respectively.
Client runs $\fn{Setup}$, signs the Merkle root and Merkle Tree height, and passes $(d,h,\fn{Sign}(sk_c,d||h))$ to the storage provider. 
The storage provider runs $\fn{Setup}$ separately, and if it gets the same result, he then signs $(d,h)$, and sends $(d,h,\fn{Sign}(sk_c,d||h), \fn{Sign}(sk_{sp},d||h))$ to the smart-contract. The smart-contract verifies signatures and stores $d$ on-chain for future proof of storage verification.

\noindent\textbf{\textit{Discussion}}

Based on our definition of PoS, several different tentative solutions can be applied.
Various cryptographic accumulators \cite{boneh2019batching, DBLP:conf/asiacrypt/GhoshOPTT16} and vector commitments \cite{catalano2013vector, lai2019subvector, boneh2019batching, campanelli2020incrementally} are valid solutions. 
However, we choose the Merkel tree to construct the PoS based on mainly two reasons. 

Firstly, in the decentralized settings, we desire no trusted setup and less public parameters which limits the use of RSA and bilinear pairing based solutions.

Secondly, the cost of the proof and verification has been considered in the incentive layer. 
The major concern of our scheme is the digest size because the digest needs to be stored on-chain.
On this point, the Merkle tree is a good solution as we only need to store the Merkle root on-chain which is a hash value.

\subsubsection{Counting the number of requests}

The number of requests for retrieving data during the storage contract is an important factor in pricing the storage service. This is due to the fact that the number of retrieval requests directly impacts the storage provider's workload. For example, consider that a client demands a service with a maximum of 5 retrieval times in a year. On the other hand, another client demands 1000 retrieval times in a year. Therefore, the storage service should be able to dynamically charge clients based on the number of reading requests. Currently, DSNs do not support the number of requests in the storage service. 

A naive approach is to apply request counting. In this case, the client first signs a request for data and sends it to the storage provider. The storage provider then verifies the signature and sends the data back to the client. However, as a malicious storage provider might refuse to send data, the client should send back a signed acknowledgment message upon successful delivery of data. On the other hand, a malicious client refuses to send back the signed acknowledgment message. To solve this issue, one simple approach is to split the data into smaller pieces and send the next portion of data upon receiving the previous message acknowledgment. Although such an approach is appropriate in the network layer with TCP protocol, in the application layer this approach is too costly as every acknowledgment message should be signed and verified. Moreover, there is no guarantee that the client sends the last message acknowledgment. 

To solve the naive approach, in our scheme we follow our proposed dynamic game-theoretic approach for challenging the storage provider. In our model, the client only sends a signed request message. The storage provider has two options, whether send or not to send the data. Then, it is the client's choice to challenge or not challenge. 
Following our dynamic game tree, the best outcome will be reached through sharing and not challenging strategies. 

For cashing out the number of requests, the storage provider only needs to submit the last signed request message. Note that the request message includes a counter indicating the number of requests so far that have been submitted by the client.

%

\section{Implementation}\label{ana}

\begin{figure*}[th]
	\captionsetup{justification=centering}
	\centering
	\begin{subfigure}[b]{0.475\linewidth}
		\centering
		\includegraphics[width=\textwidth]{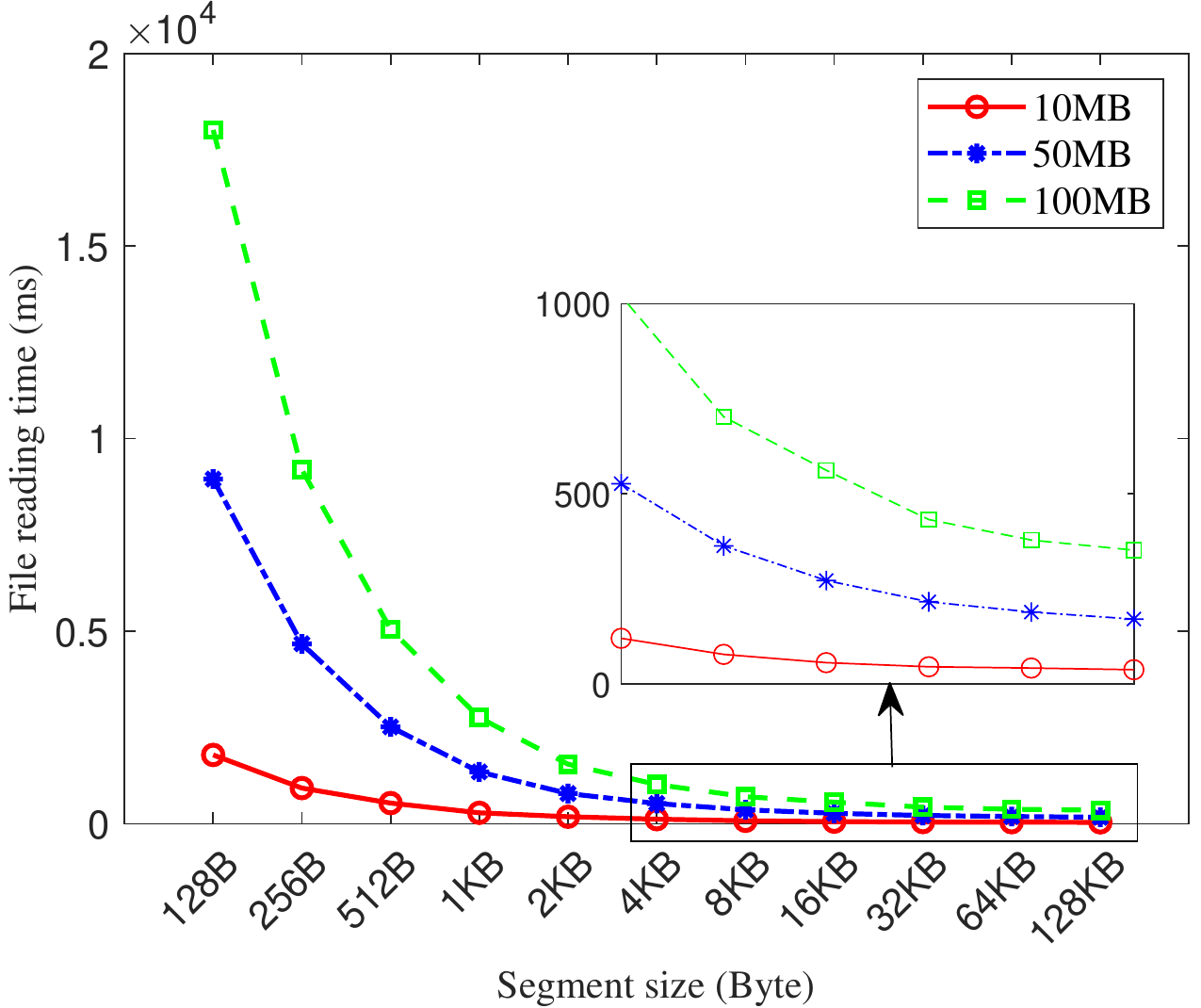}
	\end{subfigure}
	\hfill
	\begin{subfigure}[b]{0.49\linewidth}
		\centering
		\includegraphics[width=\textwidth]{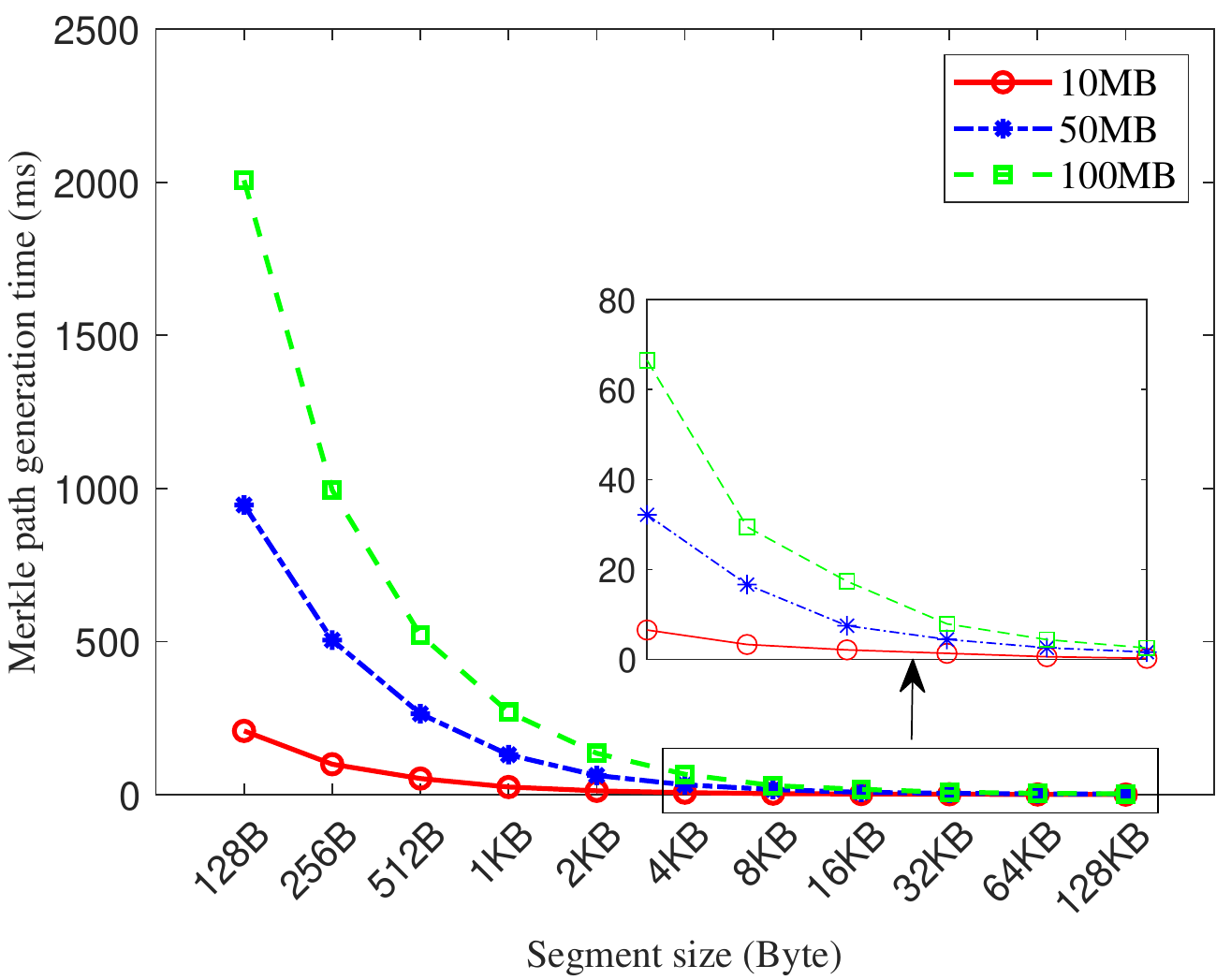}
	\end{subfigure}
	\hfill
	\begin{subfigure}[b]{0.49\linewidth}
		\centering
		\includegraphics[width=\textwidth]{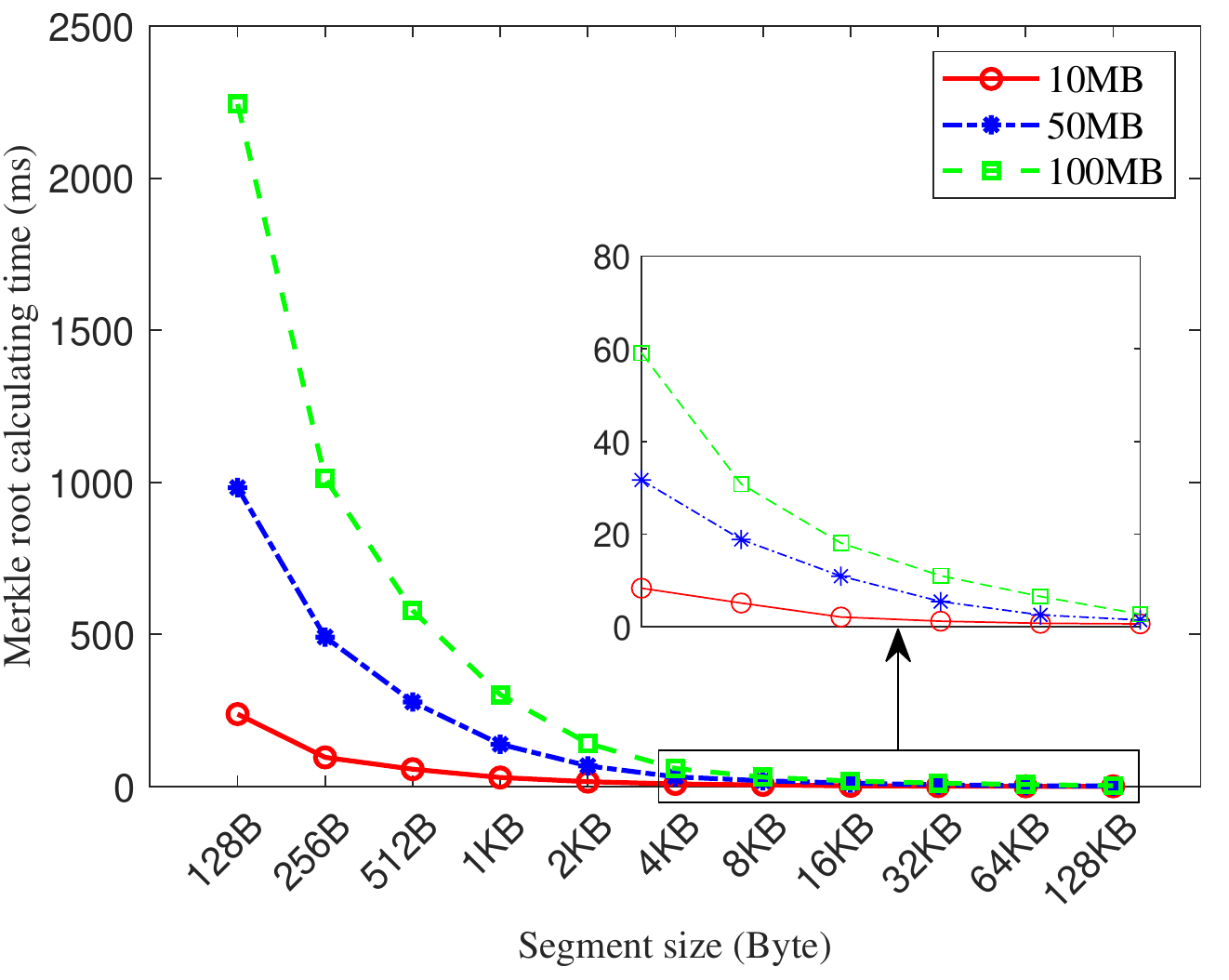}
	\end{subfigure}
	\hfill
	\begin{subfigure}[b]{0.49\linewidth}
		\centering
		\includegraphics[width=\textwidth]{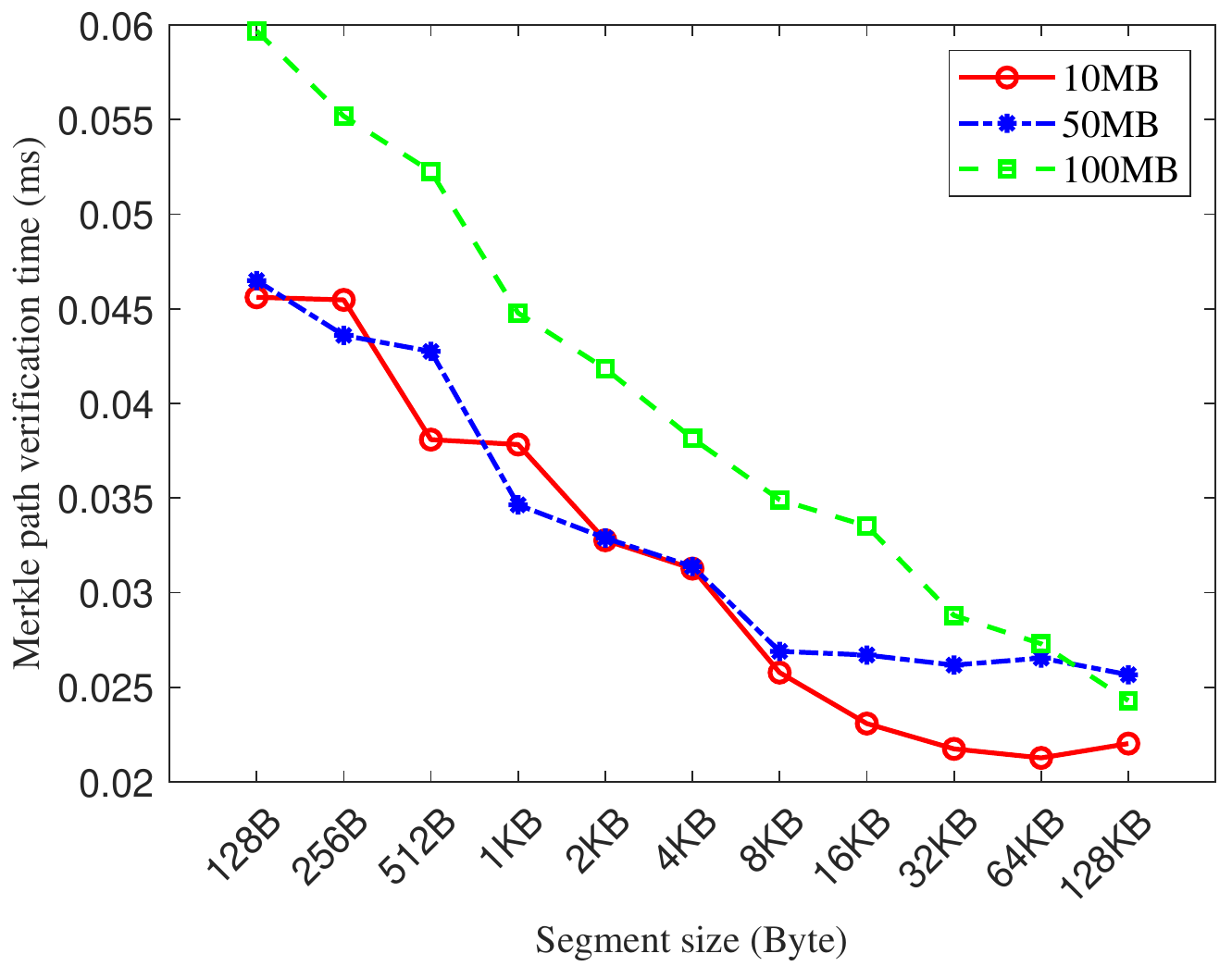}
	\end{subfigure}
	\hfill
	\caption{Computation cost of File reading, Merkle path generation, merkle root calculation, and Merkle path verification for files of 10MB, 50MB, and 100MB with varying segment size}\label{fig:111}
\end{figure*}

\begin{figure*}[h]
	\centering
	\captionsetup{justification=centering}
	\begin{subfigure}[b]{0.49\linewidth}
		\centering
		\includegraphics[width=\textwidth]{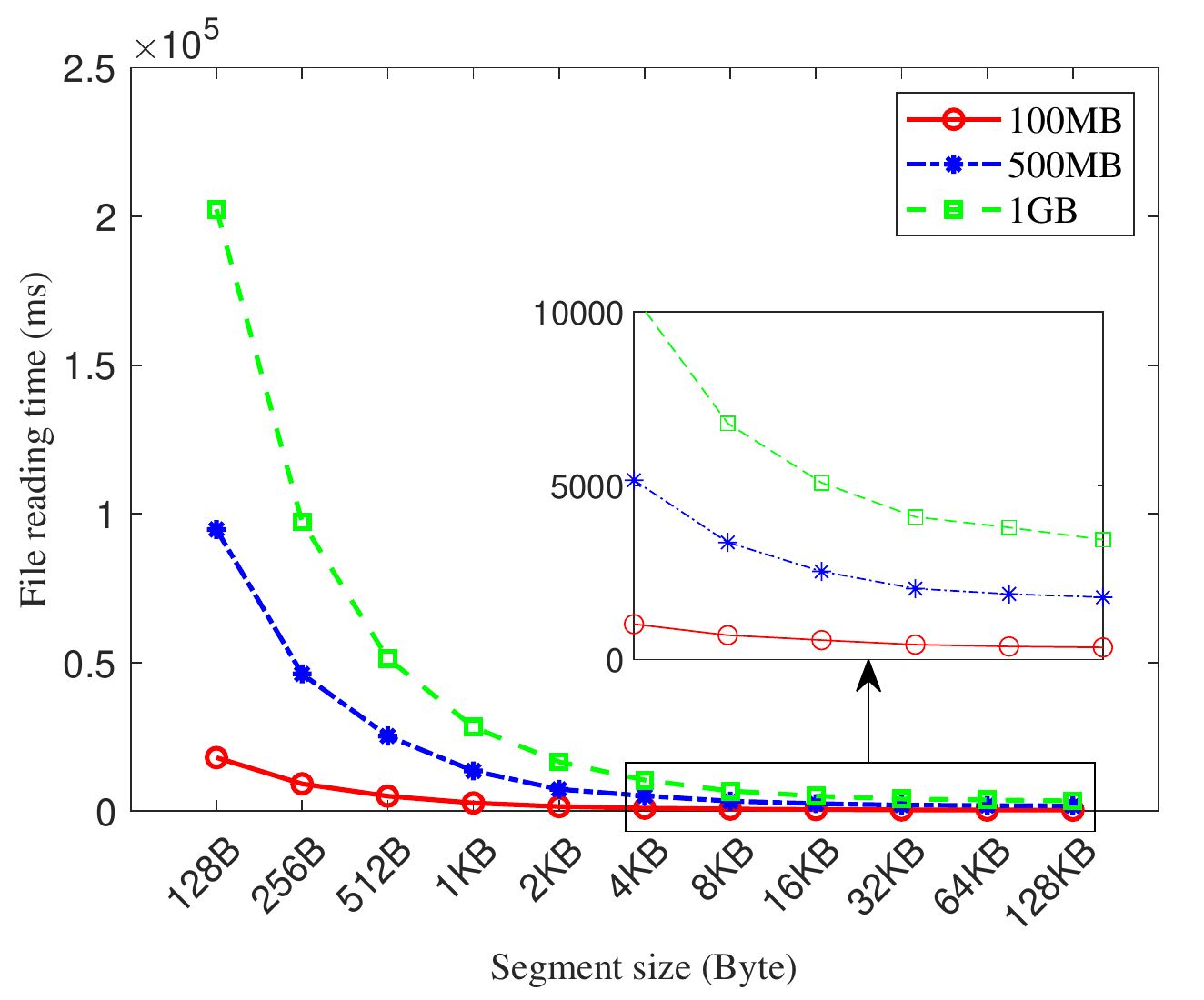}
	\end{subfigure}
	\hfill
	\begin{subfigure}[b]{0.49\linewidth}
		\centering
		\includegraphics[width=\textwidth]{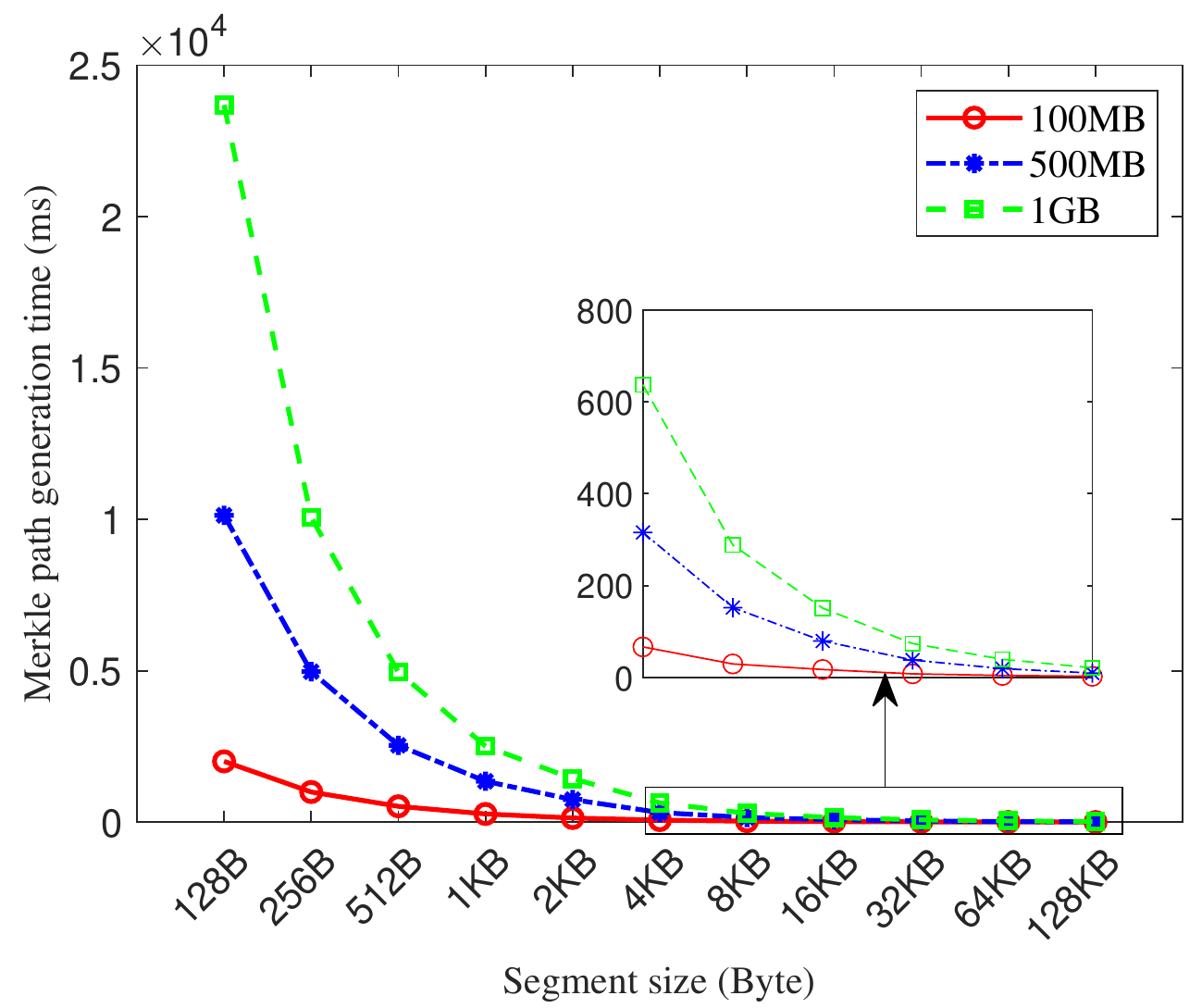}
	\end{subfigure}
	\hfill
	\begin{subfigure}[b]{0.49\linewidth}
		\centering
		\includegraphics[width=\textwidth]{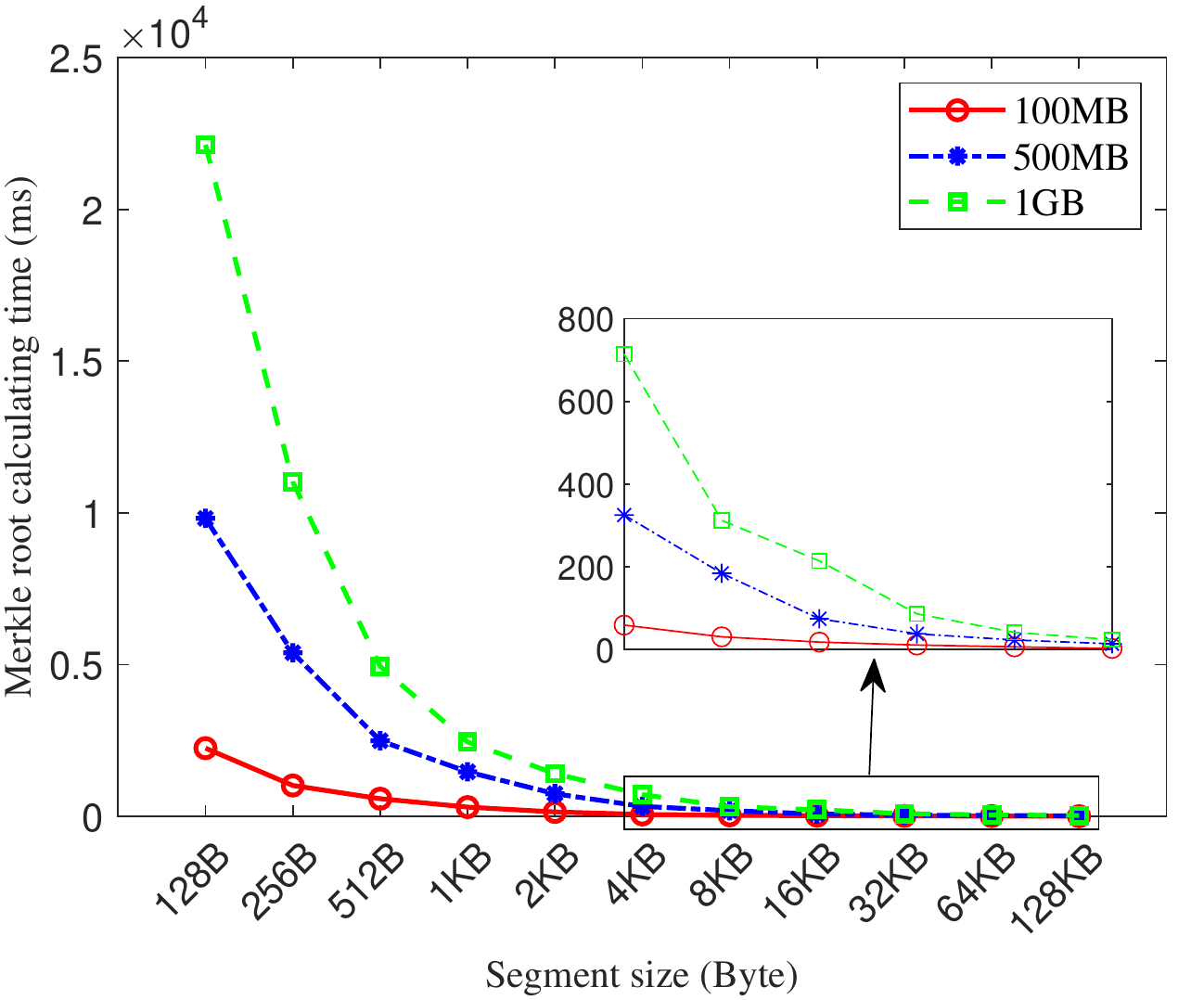}
	\end{subfigure}
	\hfill
	\begin{subfigure}[b]{0.49\linewidth}
		\centering
		\includegraphics[width=\textwidth]{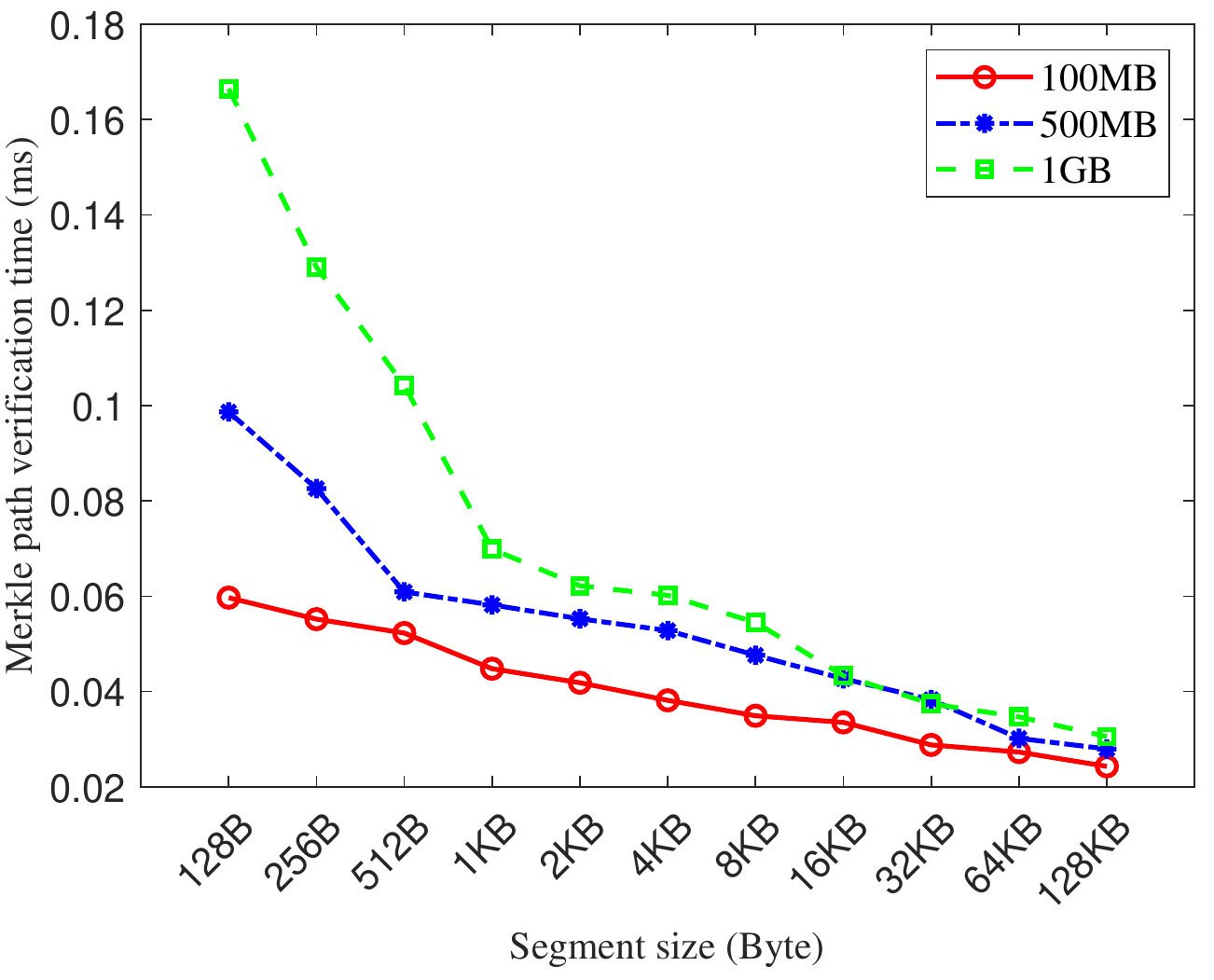}
	\end{subfigure}
	\hfill
	\caption{Computation cost of File reading, Merkle path generation, merkle root calculation, and Merkle path verification for files of 100MB, 500MB, and 1GB with varying segment size}\label{fig:222}
\end{figure*}

In this section, first, we describe the details of our implementation, and then we discuss the performance evaluation of our proposed mechanism. 
We use the Solidity (version 0.8.7) programming language to implement our smart contract. 
For the smart contract development, we use Kovan as our Blockchain solution which is an Ethereum test network allowing for blockchain development testing. We use Remix IDE to develop, deploy and administer the smart contract. 
In our implementation, we use Chainlink's oracle~\cite{chainlink} which is currently the most popular oracle network and has the majority share of the oracle market~\cite{kaleem2021demystifying}.

Considering gas fees and expensive on-chain calculation costs, we only record the basic information and the Merkle root value of the file on-chain. The storing of the original outsourced data and the verification process are done off-chain. Thus, we need to build a connection with the off-chain data. Every time the client makes an on-chain challenge request, the Chainlink core will route the assignment to an external adapter which performs a request to the API provided by the storage provider. The storage provider computes the Merkle path of the challenging piece. The external adapter processes the response using the Merkle path to calculate the root value and compares it to the on-chain stored value. Afterward, the external adapter passes the verification result back to the core. The core reports the result to the CHAINLINK-SC which in turn passes the result on-chain.

We use the EthBool core adapter provided by Chainlink to convert data to Solidity's format. We also set up a Chainlink external adapter to complete custom computations. The external adapter is written in JavaScript programming language and runs as an HTTP Server in Node.js (version 16.13.1). 
We simulate the storage provider as an HTTP server in Node.js (version 16.13.1). It provides APIs for calculating Merkle root, accessing the original file, and generating Merkle path. We divide the file with the segment size chosen by the client. The hash values of the slices are the leaf nodes of the Merkle tree. Therefore, we can build the Merkle tree and calculate the Merkle root.
For demonstration, we also developed a website that provides the clients with basic interaction with our system. The client can upload a file for storage to the storage provider, record the Merkle root on the blockchain and challenge the storage provider. We use the React library to develop the front-end application and use Parcel as the building tool.

The implementation of our proposed scheme is accessible to the public through the following GitHub repository link:

\texttt{https://github.com/podiumdesu/ICM-DSN}

The cost of storage contract deployment, recording a storage task, and submitting a challenge request is depicted in Table~\ref{gas}. 
Note that once the storage contract is deployed, different storage tasks can be recorded on top of it. Moreover, to decrease the cost of on-chain deployment, a pooling approaches can be applied~\cite{vakilinia2019pooling} without affecting the integrity of the scheme.

\begin{table}[bh]
	\caption{Gas cost for the storage contract}
	\vspace{-.1in}
	\begin{center}
		\begin{tabular}{ |c|c|c| } 
			\hline
			\rowcolor{lightgray} Operation & Gas units \\ [1ex]
			\hline
			Contract on-chain Deployment & 2,491,606   \\
			\hline 
			Recording a Storage Task & 202,001 \\ 
			\hline
			Challenge Request & 192,101  \\ 
			\hline
		\end{tabular}
		\label{gas}
	\end{center}
	\vspace{-.15in}
\end{table}


\subsection{Performance analysis}
In the subsequent experiments, we have used a macOS (version 12.0.1) laptop with an Apple M1 Pro CPU and 32 GB of memory for the performance analysis of our scheme. We conducted our evaluation over files with various sizes of 10MB, 50MB, 100MB, 500MB, and 1GB. The computing time is calculated with varying segment sizes for the files. We included various settings for our evaluation. Specifically, we look at comparisons between four dimensions of file reading time, Merkle root calculating time, Merkle path generation time, and Merkle path verification time. Figures~\ref{fig:111} demonstrate the computation cost of file reading time, Merkle rot calculation, Merkle path generation, and Merkle path verification for files with sizes of 10MB, 50MB, and 100MB.
Figures~\ref{fig:222} demonstrates the same settings for files with sizes of 100MB, 500MB, and 1GB.

As can be seen, with the increase in segment size, the computation cost for file reading time, Merkle root calculation, and Merkle path generation is decreasing at a decreasing rate. Merkle path verification is fast, and its time is also decreasing with the increase of the segment size. 

When the file sizes of the same segment size increase, the time of file reading, Merkle root calculating and Merkle path generation has the same growth. In contrast, the time of Merkle path verification time is almost the same at a very small value. 
The detailed values of our experimental results are presented in Appendix 1. 

\section{Conclusion}\label{con}

In this paper, we have introduced a novel game-theoretic mechanism for the decentralized storage network allowing the client to challenge the storage provider. This allows us to eliminate the requirement of having continuously verifying the storage provider which in turn improves the performance of DSNs. Moreover, the client is protected from service denying attack where a dishonest storage provider submits proof of storage to the network while refusing service to the client.
Our proposed model is pluggable into any blockchain platform with smart contract execution capability.
We leverage the smart contract and oracle network to govern the rules of the storage contract. We have implemented our scheme using Solidity language and Chainlink oracle network. The performance result demonstrates the applicability of our scheme.

\appendices 
\bibliographystyle{ieeetran}
\bibliography{ref.bib}

\begin{thebibliography}{10}

\bibitem{de2021exploring}
S.~de~Figueiredo, A.~Madhusudan, V.~Reniers, S.~Nikova, and B.~Preneel,
  ``Exploring the storj network: a security analysis,'' in {\em Proceedings of
  the 36th Annual ACM Symposium on Applied Computing}, pp.~257--264, 2021.

\bibitem{ateniese2020proof}
G.~Ateniese, L.~Chen, M.~Etemad, and Q.~Tang, ``Proof of storage-time:
  Efficiently checking continuous data availability,'' in {\em Proceedings of
  the 25th network and distributed system security symposium (NDSS)},
  pp.~1--15, Internet Society, 2020.

\bibitem{filecoin}
``Filecoin: A decentralized storage network,'' tech. rep., Protocol Labs, 2017.

\bibitem{sia}
D.~Vorick and L.~Champine, ``Sia: Simple decentralized storage,'' tech. rep.,
  2014.

\bibitem{storj}
``Storj: A decentralized cloud storage network framework,'' tech. rep., Storj
  Labs, Inc., 2018.

\bibitem{ethersphere2016sw3}
V.~Tron, A.~Fischer, D.~N. A, Z.~Felföldi, and N.~Johnson, ``swap, swear and
  swindle: incentive system for swarm,'' tech. rep., Ethersphere, 2016.
\newblock Ethersphere Orange Papers 1.

\bibitem{ipfs}
IPFS, ``Interplanetary file system.''

\bibitem{dziembowski2015proofs}
S.~Dziembowski, S.~Faust, V.~Kolmogorov, and K.~Pietrzak, ``Proofs of space,''
  in {\em Advances in Cryptology - {CRYPTO} 2015 - 35th Annual Cryptology
  Conference, Santa Barbara, CA, USA, August 16-20, 2015, Proceedings, Part
  {II}} (R.~Gennaro and M.~Robshaw, eds.), vol.~9216 of {\em Lecture Notes in
  Computer Science}, pp.~585--605, Springer, 2015.

\bibitem{ateniese2007provable}
G.~Ateniese, R.~Burns, R.~Curtmola, J.~Herring, L.~Kissner, Z.~Peterson, and
  D.~Song, ``Provable data possession at untrusted stores,'' in {\em
  Proceedings of the 14th ACM conference on Computer and communications
  security}, pp.~598--609, 2007.

\bibitem{ateniese2008scalable}
G.~Ateniese, R.~Di~Pietro, L.~V. Mancini, and G.~Tsudik, ``Scalable and
  efficient provable data possession,'' in {\em Proceedings of the 4th
  international conference on Security and privacy in communication netowrks},
  pp.~1--10, 2008.

\bibitem{erway2015dynamic}
C.~C. Erway, A.~K{\"u}p{\c{c}}{\"u}, C.~Papamanthou, and R.~Tamassia, ``Dynamic
  provable data possession,'' {\em ACM Transactions on Information and System
  Security (TISSEC)}, vol.~17, no.~4, pp.~1--29, 2015.

\bibitem{ben2013snarks}
E.~Ben-Sasson, A.~Chiesa, D.~Genkin, E.~Tromer, and M.~Virza, ``Snarks for c:
  Verifying program executions succinctly and in zero knowledge,'' in {\em
  Annual cryptology conference}, pp.~90--108, Springer, 2013.

\bibitem{campanelli2020incrementally}
M.~Campanelli, D.~Fiore, N.~Greco, D.~Kolonelos, and L.~Nizzardo,
  ``Incrementally aggregatable vector commitments and applications to
  verifiable decentralized storage,'' in {\em International Conference on the
  Theory and Application of Cryptology and Information Security}, pp.~3--35,
  Springer, 2020.

\bibitem{du2021enabling}
Y.~Du, H.~Duan, A.~Zhou, C.~Wang, M.~H. Au, and Q.~Wang, ``Enabling secure and
  efficient decentralized storage auditing with blockchain,'' {\em IEEE
  Transactions on Dependable and Secure Computing}, 2021.

\bibitem{yu2021efficient}
H.~Yu, Q.~Hu, Z.~Yang, and H.~Liu, ``Efficient continuous big data integrity
  checking for decentralized storage,'' {\em IEEE Transactions on Network
  Science and Engineering}, vol.~8, no.~2, pp.~1658--1673, 2021.

\bibitem{catalano2013vector}
D.~Catalano and D.~Fiore, ``Vector commitments and their applications,'' in
  {\em International Workshop on Public Key Cryptography}, pp.~55--72,
  Springer, 2013.

\bibitem{lai2019subvector}
R.~W. Lai and G.~Malavolta, ``Subvector commitments with application to
  succinct arguments,'' in {\em Annual International Cryptology Conference},
  pp.~530--560, Springer, 2019.

\bibitem{boneh2019batching}
D.~Boneh, B.~B{\"u}nz, and B.~Fisch, ``Batching techniques for accumulators
  with applications to iops and stateless blockchains,'' in {\em Annual
  International Cryptology Conference}, pp.~561--586, Springer, 2019.

\bibitem{buchmann1988key}
J.~Buchmann and H.~C. Williams, ``A key-exchange system based on imaginary
  quadratic fields,'' {\em Journal of Cryptology}, vol.~1, no.~2, pp.~107--118,
  1988.

\bibitem{libert2010concise}
B.~Libert and M.~Yung, ``Concise mercurial vector commitments and independent
  zero-knowledge sets with short proofs,'' in {\em Theory of Cryptography
  Conference}, pp.~499--517, Springer, 2010.

\bibitem{gorbunov2020pointproofs}
S.~Gorbunov, L.~Reyzin, H.~Wee, and Z.~Zhang, ``Pointproofs: Aggregating proofs
  for multiple vector commitments,'' in {\em Proceedings of the 2020 ACM SIGSAC
  Conference on Computer and Communications Security}, pp.~2007--2023, 2020.

\bibitem{srinivasan2021hyperproofs}
S.~Srinivasan, A.~Chepurnoy, C.~Papamanthou, A.~Tomescu, and Y.~Zhang,
  ``Hyperproofs: Aggregating and maintaining proofs in vector commitments,''
  {\em Cryptology ePrint Archive}, 2021.

\bibitem{papamanthou2013streaming}
C.~Papamanthou, E.~Shi, R.~Tamassia, and K.~Yi, ``Streaming authenticated data
  structures,'' in {\em Annual International Conference on the Theory and
  Applications of Cryptographic Techniques}, pp.~353--370, Springer, 2013.

\bibitem{shacham2008compact}
H.~Shacham and B.~Waters, ``Compact proofs of retrievability,'' in {\em
  International Conference on the Theory and Application of Cryptology and
  Information Security}, pp.~90--107, Springer, 2008.

\bibitem{juels2007pors}
A.~Juels and B.~S. Kaliski~Jr, ``Pors: Proofs of retrievability for large
  files,'' in {\em Proceedings of the 14th ACM conference on Computer and
  communications security}, pp.~584--597, Acm, 2007.

\bibitem{bowers2009proofs}
K.~D. Bowers, A.~Juels, and A.~Oprea, ``Proofs of retrievability: Theory and
  implementation,'' in {\em Proceedings of the 2009 ACM workshop on Cloud
  computing security}, pp.~43--54, 2009.

\bibitem{cash2017dynamic}
D.~Cash, A.~K{\"u}p{\c{c}}{\"u}, and D.~Wichs, ``Dynamic proofs of
  retrievability via oblivious ram,'' {\em Journal of Cryptology}, vol.~30,
  no.~1, pp.~22--57, 2017.

\bibitem{dodis2009proofs}
Y.~Dodis, S.~Vadhan, and D.~Wichs, ``Proofs of retrievability via hardness
  amplification,'' in {\em Theory of Cryptography Conference}, pp.~109--127,
  Springer, 2009.

\bibitem{stefanov2012iris}
E.~Stefanov, M.~van Dijk, A.~Juels, and A.~Oprea, ``Iris: A scalable cloud file
  system with efficient integrity checks,'' in {\em Proceedings of the 28th
  Annual Computer Security Applications Conference}, pp.~229--238, 2012.

\bibitem{shi2013practical}
E.~Shi, E.~Stefanov, and C.~Papamanthou, ``Practical dynamic proofs of
  retrievability,'' in {\em Proceedings of the 2013 ACM SIGSAC conference on
  Computer \& communications security}, pp.~325--336, 2013.

\bibitem{fisch2018scaling}
B.~Fisch, J.~Bonneau, N.~Greco, and J.~Benet, ``Scaling proof-of-replication
  for filecoin mining,'' tech. rep., Technical report, Stanford University,
  2018. https://web. stanford. edu~…, 2018.

\bibitem{oracle}
Chainlink, ``What is the blockchain oracle problem?,'' tech. rep., 2020.

\bibitem{egberts2017oracle}
A.~Egberts, ``The oracle problem-an analysis of how blockchain oracles
  undermine the advantages of decentralized ledger systems,'' {\em Available at
  SSRN 3382343}, 2017.

\bibitem{katz2020introduction}
J.~Katz and Y.~Lindell, {\em Introduction to modern cryptography}.
\newblock CRC press, 2020.

\bibitem{DBLP:conf/asiacrypt/GhoshOPTT16}
E.~Ghosh, O.~Ohrimenko, D.~Papadopoulos, R.~Tamassia, and N.~Triandopoulos,
  ``Zero-knowledge accumulators and set algebra,'' in {\em Advances in
  Cryptology - {ASIACRYPT} 2016 - 22nd International Conference on the Theory
  and Application of Cryptology and Information Security, Hanoi, Vietnam,
  December 4-8, 2016, Proceedings, Part {II}} (J.~H. Cheon and T.~Takagi,
  eds.), vol.~10032 of {\em Lecture Notes in Computer Science}, pp.~67--100,
  2016.

\bibitem{chainlink}
``Chainlink 2.0: Next steps in the evolution of decentralized storage
  networks,'' tech. rep., Chainlink Labs, 2021.

\bibitem{kaleem2021demystifying}
M.~Kaleem and W.~Shi, ``Demystifying pythia: A survey of chainlink oracles
  usage on ethereum,'' in {\em International Conference on Financial
  Cryptography and Data Security}, pp.~115--123, Springer, 2021.

\bibitem{vakilinia2019pooling}
I.~Vakilinia, S.~Vakilinia, S.~Badsha, E.~Arslan, and S.~Sengupta, ``Pooling
  approach for task allocation in the blockchain based decentralized storage
  network,'' in {\em 2019 15th International Conference on Network and Service
  Management (CNSM)}, pp.~1--6, IEEE, 2019.

\end{thebibliography}


\appendix

\noindent\textbf{Appendix A.}

The detailed computation cost for our performance analysis on files with different sizes is presented in the following tables. 

\begin{table*}[hbt!]
	\caption{File with the size of 10MB }
	\vspace{-.1in}
	\begin{center}
		\begin{tabular}{ |m{2cm}|m{2cm}|m{2cm}|m{2cm}|m{2cm}|m{2cm}|m{2cm}| } 
			\hline
			\rowcolor{lightgray}  Segment Size & Number of Segments & Tree Height & File Reading Time (ms) & Merkle Root Calculating Time (ms) & Merkle Path Generating Time (ms) & Merkle Path Verification Time (ms)\\ [1ex]
			\hline
			128 bytes & 81920 & 17 & 1787.904 & 238.09 & 208.206 & 0.04561 \\
			\hline
			256 bytes & 40960 & 16 & 927.353 & 95.722 & 99.334 & 0.045473 \\
			\hline
			512 bytes & 20480 & 15 & 537.782 & 56.897 & 52.63 & 0.038086 \\
			\hline
			1024 bytes & 10240 & 14 & 285.219 & 29.52 & 24.91 & 0.037828 \\
			\hline
			2048 bytes & 5120 & 13 & 183.741 & 15.773 & 12.442 & 0.032765 \\
			\hline
			4KB & 2560 & 12 & 120.084 & 8.305 & 6.529 & 0.031265 \\
			\hline
			8KB & 1280 & 11 & 77.793 & 5.104 & 3.31 & 0.025769 \\
			\hline
			16KB & 640 & 10 & 56.051 & 2.083 & 2.108 & 0.023075 \\
			\hline
			32KB & 320 & 9 & 45.227 & 1.185 & 1.342 & 0.021736 \\
			\hline
			64KB & 160 & 8 & 41.836 & 0.725 & 0.587 & 0.021259 \\
			\hline
			128KB & 80 & 7 & 37.5 & 0.605 & 0.237 & 0.022016 \\
			\hline
		\end{tabular}
		\label{10}
	\end{center}
	\vspace{-.15in}
\end{table*}

\begin{table*}[h]
	\caption{File with the size of 50MB}
	\vspace{-.1in}
	\begin{center}
		\begin{tabular}{ |m{2cm}|m{2cm}|m{2cm}|m{2cm}|m{2cm}|m{2cm}|m{2cm}| } 
			\hline
			\rowcolor{lightgray}  Segment Size & Number of Segments & Tree Height & File Reading Time (ms) & Merkle Root Calculating Time (ms) & Merkle Path Generating Time (ms) & Merkle Path Verification Time (ms)\\ [1ex]
			\hline
			128 bytes & 409600 & 19 & 8950.15 & 982.823 & 946.094 & 0.046496 \\
			\hline
			256 bytes & 204800 & 18 & 4672.522 & 491.322 & 504.713 & 0.043598 \\
			\hline
			512 bytes & 102400 & 17 & 2515.876 & 277.763 & 264.181 & 0.042748 \\
			\hline
			1024 bytes & 51200 & 16 & 1347.252 & 138.973 & 130.389 & 0.034638 \\
			\hline
			2048 bytes & 25600 & 15 & 786.049 & 67.828 & 62.617 & 0.032882 \\
			\hline
			4KB & 12800 & 14 & 526.626 & 31.673 & 32.167 & 0.031378 \\
			\hline
			8KB & 6400 & 13 & 363.279 & 18.832 & 16.647 & 0.026893 \\
			\hline
			16KB & 3200 & 12 & 272.149 & 10.909 & 7.483 & 0.026698 \\
			\hline
			32KB & 1600 & 11 & 216.035 & 5.435 & 4.442 & 0.026167 \\
			\hline
			64KB & 800 & 10 & 189.12 & 2.535 & 2.575 & 0.026538 \\
			\hline
			128KB & 400 & 9 & 170.561 & 1.441 & 1.62 & 0.025657 \\
			\hline
		\end{tabular}
		\label{50}
	\end{center}
	\vspace{-.15in}
\end{table*}

\begin{table*}[h]
	\caption{File with the size of 100MB}
	\vspace{-.1in}
	\begin{center}
		\begin{tabular}{ |m{2cm}|m{2cm}|m{2cm}|m{2cm}|m{2cm}|m{2cm}|m{2cm}| } 
			\hline
			\rowcolor{lightgray}  Segment Size & Number of Segments & Tree Height & File Reading Time (ms) & Merkle Root Calculating Time (ms) & Merkle Path Generating Time (ms) & Merkle Path Verification Time (ms)\\ [1ex]
			\hline
			128 bytes & 819200 & 20 & 18014.235 & 2244.158 & 2006.815 & 0.059693 \\
			\hline
			256 bytes & 409600 & 19 & 9195.558 & 1013.982 & 994.447 & 0.055185 \\
			\hline
			512 bytes & 204800 & 18 & 5049.277 & 578.276 & 521.768 & 0.052264 \\
			\hline
			1KB & 102400 & 17 & 2758.359 & 301.883 & 269.799 & 0.044779 \\
			\hline
			2KB & 51200 & 16 & 1547.833 & 141.994 & 135.149 & 0.041836 \\
			\hline
			4KB & 25600 & 15 & 1020.346 & 59.108 & 66.528 & 0.038147 \\
			\hline
			8KB & 12800 & 14 & 702.577 & 30.809 & 29.473 & 0.034908 \\
			\hline
			16KB & 6400 & 13 & 562.009 & 18.067 & 17.364 & 0.03352 \\
			\hline
			32KB & 3200 & 12 & 432.415 & 10.992 & 7.898 & 0.028791 \\
			\hline
			64KB & 1600 & 11 & 378.376 & 6.522 & 4.389 & 0.027291 \\
			\hline
			128KB & 800 & 10 & 351.862 & 2.69 & 2.538 & 0.024287 \\
			\hline
		\end{tabular}
		\label{100}
	\end{center}
	\vspace{-.15in}
\end{table*}

\begin{table*}[h]
	\caption{File with the size of 500MB}
	\vspace{-.1in}
	\begin{center}
		\begin{tabular}{ |m{2cm}|m{2cm}|m{2cm}|m{2cm}|m{2cm}|m{2cm}|m{2cm}| } 
			\hline
			\rowcolor{lightgray}  Segment Size & Number of Segments & Tree Height & File Reading Time (ms) & Merkle Root Calculating Time (ms) & Merkle Path Generating Time (ms) & Merkle Path Verification Time (ms)\\ [1ex]
			\hline
			128 bytes & 4096000 & 22 & 94737.491 & 9818.019 & 10132.573 & 0.098665 \\
			\hline
			256 bytes & 2048000 & 21 & 46169.824 & 5388.3 & 4969.759 & 0.082575 \\
			\hline
			512 bytes & 1024000 & 20 & 25267.874 & 2489.126 & 2534.864 & 0.060895 \\
			\hline
			1KB & 512000 & 19 & 13553.087 & 1464.32 & 1348.817 & 0.058168 \\
			\hline
			2KB & 256000 & 18 & 7354.601 & 741.567 & 743.942 & 0.055256 \\
			\hline
			4KB & 128000 & 17 & 5156.859 & 325.235 & 315.587 & 0.052793 \\
			\hline
			8KB & 64000 & 16 & 3368.927 & 184.248 & 152.093 & 0.047618 \\
			\hline
			16KB & 32000 & 15 & 2534.908 & 74.448 & 79.405 & 0.042676 \\
			\hline
			32KB & 16000 & 14 & 2041.117 & 38.056 & 37.963 & 0.038281 \\
			\hline
			64KB & 8000 & 13 & 1886.029 & 23.166 & 19.252 & 0.030165 \\
			\hline
			128KB & 4000 & 12 & 1792.042 & 13.81 & 9.572 & 0.027942 \\
			\hline
		\end{tabular}
		\label{500}
	\end{center}
	\vspace{-.15in}
\end{table*}

\begin{table*}[h]
	\caption{File with the size of 1GB}
	\vspace{-.1in}
	\begin{center}
		\begin{tabular}{ |m{2cm}|m{2cm}|m{2cm}|m{2cm}|m{2cm}|m{2cm}|m{2cm}| } 
			\hline
			\rowcolor{lightgray}  Segment Size & Number of Segments & Tree Height & File Reading Time (ms) & Merkle Root Calculating Time (ms) & Merkle Path Generating Time (ms) & Merkle Path Verification Time (ms)\\ [1ex]
			\hline
			128 bytes & 8192000 & 23 & 202395.99 & 22132.426 & 23687.559 & 0.166503 \\
			\hline
			256 bytes & 4096000 & 22 & 97338.305 & 11019.89 & 10064.421 & 0.129042 \\
			\hline
			512 bytes & 2048000 & 21 & 51409.273 & 4939.284 & 4955.69 & 0.104241 \\
			\hline
			1KB & 1024000 & 20 & 28270.954 & 2449.939 & 2510.657 & 0.069888 \\
			\hline
			2KB & 512000 & 19 & 16586.863 & 1404.678 & 1435.893 & 0.062182 \\
			\hline
			4KB & 256000 & 18 & 10458.671 & 714.115 & 636.902 & 0.060185 \\
			\hline
			8KB & 128000 & 17 & 6790.474 & 312.334 & 288.44 & 0.054482 \\
			\hline
			16KB & 64000 & 16 & 5085.178 & 214.378 & 150.928 & 0.043327 \\
			\hline
			32KB & 32000 & 15 & 4101.697 & 86.309 & 73.861 & 0.03741 \\
			\hline
			64KB & 16000 & 14 & 3798.152 & 41.595 & 39.496 & 0.034669 \\
			\hline
			128KB & 8000 & 13 & 3446.469 & 23.972 & 20.959 & 0.030534 \\
			\hline
		\end{tabular}
		\label{1GB}
	\end{center}
	\vspace{-.15in}
\end{table*}

\end{document}